\documentclass[%
reprint,
superscriptaddress,
 amsmath,amssymb,
 aps,
 prc,
longbibliography]{revtex4-1}

\usepackage{graphicx}% Include figure files
\usepackage{mathtools}
\usepackage{subfigure}
\usepackage{dcolumn}% Align table columns on decimal point
\usepackage{bm}% bold math
\usepackage{epstopdf}
\usepackage{xcolor}
\usepackage{hyperref} % add hypertext capabilities
\usepackage{appendix} % ADDED 6/13/2019
\usepackage{float} % \usepackage[mathlines]{lineno}
\begin{document}

% \preprint{APS/123-QED}
\title{Resonances and collisional properties of neutron-rich helium isotopes in the adiabatic hyperspherical representation}
%\title{2024 Nuclear paper on Helium Isotopes using the adiabatic hyperspherical representation: version 2}
% Force line breaks with \\
% \thanks{A footnote to the article title} %

\author{Michael D. Higgins}
\affiliation{Department of Physics and Astronomy, Purdue University, West Lafayette, Indiana 47907 USA}
\author{Chris H. Greene}
\affiliation{Department of Physics and Astronomy, Purdue University, West Lafayette, Indiana 47907 USA}
\affiliation{Purdue Quantum Science and Engineering Institute, Purdue University, West Lafayette, Indiana 47907 USA}
 %\altaffiliation[Also at ]{Physics Department, XYZ University.}%Lines break automatically or can be forced with \\

\date{\today}% It is always \today, today,
             %  but any date may be explicitly specified

\begin{abstract}
This work treats few--body systems consisting of neutrons interacting with a $\prescript{4}{}{\mathrm{He}}$ nucleus. The adiabatic hyperspherical representation is utilized to solve the $N$--body Schr$\ddot{\mathrm{o}}$dinger equation for the three-- and four--body systems, treating both $\prescript{6}{}{\mathrm{He}}$ and $\prescript{7}{}{\mathrm{He}}$ nuclei. A simplified central potential model for the $\prescript{4}{}{\mathrm{He}}-n$ interaction is used in conjunction with a spin--dependent three--body interaction to reproduce $\prescript{6}{}{\mathrm{He}}$ bound--state and resonance properties as well as properties for the $\prescript{8}{}{\mathrm{He}}$ nucleus in its ground--state. With this Hamiltonian, the adiabatic hyperspherical representation is used to compute bound and scattering states for both $\prescript{6}{}{\mathrm{He}}$ and $\prescript{7}{}{\mathrm{He}}$ nuclei. For the $\prescript{6}{}{\mathrm{He}}$ system, the electric quadrupole transition between the $0^{+}$ and $2^{+}$ state is investigated. For the $\prescript{7}{}{\mathrm{He}}$ system, $\prescript{6}{}{\mathrm{He}}+n$ elastic scattering is investigated along with the four--body recombination process $\prescript{4}{}{\mathrm{He}}+n+n+n\rightarrow\prescript{6}{}{\mathrm{He}}+n$ and breakup process $\prescript{6}{}{\mathrm{He}}+n\rightarrow\prescript{4}{}{\mathrm{He}}+n+n+n$.
%An area of the few-body sector that provides a natural class of systems near the unitarity limit is low-energy nuclear physics. These systems include halo nuclei, which are good candidate systems to study the near-unitarity limit and universality. Two–neutron halos have been extensively studied, such as $\prescript{6}{}{\mathrm{He}}$, 9B, 11Li, 12Be, and 12C. With the large neutron-neutron ($nn$) $^{1}S_{0}$ scattering length ($a_{s}\approx-18.5$ fm), halo nuclei are good candidates to study universal physics near the unitarity limit and the possible emergence of Efimov physics in Borromean systems or the scattering continuum. In this work, we use the adiabatic hyperspherical representation to study halo nuclei consisting of four neutrons. This work focuses on the 4He+2n , 4He+3n , and 4He+4n systems. In particular, the 4He+4n system has been of recent theoretical interest sparked by the 2022 experiment by Duer et. al. that found a low-energy 4n signal in the missing mass spectrum of the 8He(p,p4He)4n reaction. This work aims to provide theoretical insight into the qualitative and quantitative nature of the five-body scattering continuum and possible universal physics arising in four-neutron halos from the nn interaction.
\end{abstract}

\pacs{Valid PACS appear here}% PACS, the Physics and Astronomy
                             % Classification Scheme.
%\keywords{Suggested keywords}%Use showkeys class option if keyword
                              %display desired
\maketitle

\section{Introduction}

Halo nuclei are interesting few--body systems to study nuclear systems near the unitarity limit. These nuclei consist of a core nucleus and nucleons, where the nucleons are, on average, far away from the core. Thus, the short–range behavior of the core–neutron interaction should not be important and thus model independence should arise. Most halo nuclei studied are two neutron halos such as $^{6}\mathrm{He}$ \cite{PhysRevLett.79.2411,Thompson1998,GARRIDO1997153,PhysRevC.90.064301,PhysRevLett.113.032503}, $^{11}\mathrm{Li}$ \cite{DANILIN1994299,Thompson1998,GARRIDO1997153}, $^{12}\mathrm{Be}$ \cite{PhysRevC.86.024310} and $^{22}\mathrm{C}$ \cite{TOGANO2016412}. With the large neutron--neutron $^{1}S_{0}$ scattering length ($a_s\approx-18.5~\mathrm{fm}$), halo nuclei are good candidates to study universal physics near the $s$--wave unitary limit and study possible emergence of Efimov physics in Borromean systems and in the scattering continuum \cite{PhysRevLett.73.2817}. One notable core--neutron system is that of $\prescript{18}{}{\mathrm{B}}$, which has an $s$--wave scattering length of $a_{s}\approx-100~\mathrm{fm}$ \cite{PhysRevC.100.011603}. This system provides a good test bed for studying Efimov character in the weakly bound Borromean state and resonances of the two--neutron halo $\prescript{19}{}{\mathrm{B}}$ nucleus \cite{PhysRevC.100.011603}.

There are notable few--body nuclear reactions with more than two neutrons that are of importance. Most notable are the reactions involving the ejection of three or four neutrons in the final state \cite{MarquesCarbonell2021}, which serve to understand how three and four neutrons interact in the scattering continuum. Some reactions that have been studied to probe these few--neutron clusters, with three neutrons in the final state \cite{PhysRevLett.14.444,REITAN1971368,PhysRevLett.36.942}, and four neutrons in the final state \cite{PhysRevC.65.044006,10.1063/1.2746575,PhysRevLett.116.052501,PhysRevC.103.044313,DuerNat2022,muzalevskii2024}. To understand the low energy four--neutron signal in the Kisamori et al. 2016 experiment \cite{PhysRevLett.116.052501} and Duer et al. 2022 experiment \cite{DuerNat2022}, the six--body reactions $\prescript{8}{}{\mathrm{He}}(p,p{\prescript{4}{}{\mathrm{He}}})4n$ and $\prescript{4}{}{\mathrm{He}}(\prescript{8}{}{\mathrm{He}},\prescript{8}{}{\mathrm{Be}})4n$ should be investigated theoretically, where $^{8}\mathrm{Be}$ is modeled as two $\prescript{4}{}{\mathrm{He}}$ nuclei \cite{BUCK1977246}. Understanding the low--energy $4n$ signature in these experiments has been a challenge, with conflicting interpretations as to the description of the ejected complex of neutrons in the final state.

Recent studies \cite{lazauskas2022,muzalevskii2024} have been able to interpret the low energy peak in the Duer et. al. experiment via a fast removal process of the $\prescript{4}{}{\mathrm{He}}$ nucleus, leaving behind a four neutron cluster. The study by Lazauskas et al. \cite{lazauskas2022} treats the fast removal of the alpha nucleus from $^{8}\mathrm{He}$, modeling the system as four neutrons interacting with an alpha nucleus. Likewise, in Muzalevskii et. al. \cite{muzalevskii2024}, they followed a similar procedure to study low energy $4n$ populations in other experiments, such as in the reaction $\prescript{2}{}{\mathrm{H}}(\prescript{8}{}{\mathrm{He}},\prescript{6}{}{\mathrm{Li}})4n$ and $\prescript{2}{}{\mathrm{H}}(\prescript{8}{}{\mathrm{He}},\prescript{3}{}{\mathrm{He}})\prescript{7}{}{\mathrm{H}}$. In both references, low--energy peaks are present in the response function, which is approximated by the overlap of the wave function of the four--neutrons interacting with the alpha nucleus and the wave function of four free interacting neutrons at some given energy with an inhomogeneous source. Furthermore, it was also shown in \cite{lazauskas2022} that the low--energy peak in Duer et al. \cite{DuerNat2022} could also be used to obtain a measurement of the neutron--neutron $^{1}S_0$ scattering length. Thus, one might extract two--body scattering properties from other low--energy signatures from knockout reaction experiments, such as in the case of $^{6}\mathrm{He}(p,p{^{4}\mathrm{He}})2n$ \cite{DuerNat2022,PhysRevC.104.024001}.

An adiabatic hyperspherical treatment of the five--body ($\prescript{4}{}{\mathrm{He}}$+4n) system would provide further theoretical insights into the complex collision process of four interacting neutrons scattered by an $\prescript{4}{}{\mathrm{He}}$ nucleus. Studying the five--body scattering process would naturally take into account the evolution of the five--body wave function as the the overall system size changes, viewed as the hyperradius in the hyperspherical formalism \cite{DELVES1958391,DELVES1960275,Rittenhouse_2011}. In addition, when treating the hyperradius as an adiabatic parameter, the different ways the $N$--body system can break apart at a given energy can be treated on an equal footing. In contrast, methods like the no--core shell model (NCSM) \cite{PhysRevC.54.2986,PhysRevC.57.3119}, which utilize bound--state approximations, the continuum shell model (CSM) \cite{PhysRevLett.94.052501} and the Gamov--shell model (GSM) \cite{PhysRevC.71.044314}, where states representing different types of continuum pathways are introduced into previous methods. The adiabatic hyperspherical representation accounts for these channels naturally, reducing the complicated scattering problem to numerically solving a truncated subset of an infinite set of one--dimensional coupled Schr$\ddot{\mathrm{o}}$dinger equations in the hyperradial coordinate. 

A qualitative understanding of the continuum scattering process can be sought by analyzing the features of the adiabatic hyperradial potential curves. Further, a quantitative analysis of the scattering continuum can be performed through solving the coupled hyperradial Shr$\mathrm{\ddot{o}}$dinger equation. In doing so, the continuum eigen--phase shifts and Wigner--Smith time delay can be computed to study possible resonance--like features in this five--body complex. Thus, eventually our goal is to treat the $^{8}\mathrm{He}$ as described above, as four neutrons interacting with an $\prescript{4}{}{\mathrm{He}}$ nucleus using the adiabatic hyperspherical representation.

This article is organized as follows. In Sec. \ref{sec:theory_methods}, a brief overview of the adiabatic hyperspherical representation is provided, which outlines the general approach to solving the $N$--body Schr$\ddot{\mathrm{o}}$dinger equation in hyperspherical coordinates. In Sec. \ref{sec:interactions}, details on the two--body and three--body nuclear interactions used in this work are given. In Sec. \ref{sec:three_body_sys}, results for the $\prescript{6}{}{\mathrm{He}}$ system are presented, where calculations of bound and resonant states are given for the $0^{+}$ and $2^{+}$ symmetries. The electric quadrupole transition matrix element is also investigated. In Sec. \ref{sec:four_body_sys}, results for the $\prescript{7}{}{\mathrm{He}}$ system are presented, where calculations of the $3/2^{-}$ resonance state are given. Elastic and inelastic cross sections and recombination rates are also presented. In Sec. \ref{sec:five_body_sys}, bound--state properties are given for the $\prescript{8}{}{\mathrm{He}}$ system in the $0^{+}$ symmetry using the model potential developed in this work. Furthermore, preliminary results on a five--body hyperspherical calculation for this system are presented. Lastly, a summary is given in Sec. \ref{sec:conclusion}.

\section{Theoretical Methods}
\label{sec:theory_methods}
The present study solves the $N$--body Hamiltonian in hyperspherical coordinates. This method initially treats the hyperradius as an adiabatic coordinate, in what is denoted the adiabatic hyperspherical representation \cite{Rittenhouse_2011}, although nonadiabatic couplings are usually incorporated in a later stage. The general form of the Hamiltonian for $N$--body systems that will be treated in this work is given by Eq. \eqref{eq:general_nbody_H} and includes two-- and three--body interactions.
\begin{equation}
\label{eq:general_nbody_H}
\hat{H}=-\sum_{i}^{N}\frac{\hbar^{2}}{2m_{i}}\nabla^{2}_{i}+\sum_{j>i}^{N}V^{(2b)}_{ij}+\sum_{k>j>i}^{N}V^{(3b)}_{ijk}
\end{equation}
For the type of Hamiltonian in Eq. \eqref{eq:general_nbody_H}, the relative and center--of--mass motion are separable. Our interest is in finding eigenvalues and eigenvectors of the Hamiltonian of the relative motion, so recasting this Hamiltonian in hyperspherical coordinates gives

\begin{equation}
\label{eq:Full_Relative_Hamiltonian}
\hat{H}-\hat{H}_{\mathrm{c.m.}}\equiv\hat{H}_{\mathrm{rel.}}=\hat{T}_{\rho}+\hat{T}_{\Omega}+V_{\mathrm{int.}}(\rho,\Omega)
\end{equation}
where
\begin{subequations}
\begin{equation}
\label{eq:kinetic_energy_rho}
\hat{T}_{\rho}=-\frac{\hbar^{2}}{2\mu}\frac{1}{\rho^{3N-4}}\frac{\partial}{\partial\rho}\rho^{3N-4}\frac{\partial}{\partial\rho}
\end{equation}
\begin{equation}
\label{eq:kinetic_energy_omega}
\hat{T}_{\Omega}=\frac{\hbar^{2}}{2\mu\rho^{2}}\Lambda^{2}(\Omega)
\end{equation}
\end{subequations}
are the hyperradial and hyperangular kinetic energy operators, respectively \cite{Avery1989Book,Rittenhouse_2011}. The function $V_{\mathrm{int.}}(\rho,\Omega)$ is the interaction potential that contains both two-- and three--body interaction terms,
\begin{equation}
\label{eq:Vint}
V_{\mathrm{int.}}(\rho,\Omega)\equiv\sum_{j>i}V_{ij}^{(\mathrm{2b})}(\rho,\Omega)+\sum_{k>j>i}V_{ijk}^{(\mathrm{3b})}(\rho,\Omega).
\end{equation}

To diagonalize the Hamiltonian in Eq. \eqref{eq:Full_Relative_Hamiltonian}, the full $N$--body wavefunction is expanded in a hyperspherical basis $\phi_{\nu}(\rho,\Omega)$,
\begin{equation}
\label{eq:wavefunction_expanded}
\Psi_{E}(\rho,\Omega)=\rho^{-\frac{3N-4}{2}}\sum_{\nu}F_{\nu}^{E}(\rho)\phi_{\nu}(\rho,\Omega)
\end{equation}
where the basis function $\phi_{\nu}(\rho,\Omega)$ are eigenfunctions of the adiabatic Hamiltonian $H_{\mathrm{ad.}}(\rho,\Omega)$, defined as
\begin{subequations}
\begin{equation}
\label{eq:eigenvalue_eq}
H_{\mathrm{ad.}}(\rho,\Omega)\phi_{\nu}(\rho,\Omega)=U_{\nu}(\rho)\phi_{\nu}(\rho,\Omega)
\end{equation}
\begin{multline}
\label{eq:adiabatic_H}
H_{\mathrm{ad.}}(\rho,\Omega)=\frac{\hbar^{2}}{2\mu \rho^{2}}\left[\Lambda^{2}(\Omega)+\frac{(3N-4)(3N-6)}{4}\right]\\+V_{\mathrm{int.}}(\rho,\Omega).
\end{multline}
\end{subequations}
By expanding the full wave function in the hyperspherical basis in Eq. \ref{eq:wavefunction_expanded}, this leads to a an infinite set of coupled hyperradial Schr$\ddot{\mathrm{o}}$dinger equations when applying the Hamiltonian onto the expanded wave function and projecting onto it the adiabatic basis functions $\phi_{\nu}(\rho,\Omega)$. For a given channel $\nu$, the corresponding coupled hyperradial Schr$\ddot{\mathrm{o}}$dinger equation is given in Eq. \ref{eq:coupled_channel_eqs} as,
\begin{multline}
\label{eq:coupled_channel_eqs}
-\frac{\hbar^{2}}{2\mu}\frac{d^{2}}{d\rho^{2}}F_{\nu}^{E}(\rho)-\frac{\hbar^2}{2\mu}\sum_{\nu^{\prime}\neq\nu}\left[Q_{\nu\nu^{\prime}}(\rho)+2P_{\nu\nu^{\prime}}(\rho)\frac{d}{d\rho}\right]F_{\nu^{\prime}}^{E}(\rho)\\+W_{\nu}(\rho)F_{\nu}^{E}(\rho)=EF_{\nu}^{E}(\rho)
\end{multline}
where $P_{\nu\nu^{\prime}}(\rho)$ and $Q_{\nu\nu^{\prime}}(\rho)$ are the first and second derivative non--adiabatic couplings between channels $\nu$ and $\nu^{\prime}$, defined in Eqs. \eqref{eq:channel_couplings_P} and \eqref{eq:channel_couplings_Q}. 
\begin{subequations}
\begin{equation}
\label{eq:channel_couplings_P}
P_{\nu\nu^{\prime}}(\rho)=\langle\phi_{\nu}|\partial_{\rho}|\phi_{\nu^{\prime}}\rangle_{\Omega}
\end{equation}
\begin{equation}
\label{eq:channel_couplings_Q}
Q_{\nu\nu^{\prime}}(\rho)=\langle\phi_{\nu}|\partial_{\rho}^{2}|\phi_{\nu^{\prime}}\rangle_{\Omega}
\end{equation}
\end{subequations}
The function $W_{\nu}(\rho)$ is defined as the effective hyperradial potential and is given in Eq. \eqref{eq:effective_potential}.
\begin{equation}
\label{eq:effective_potential}
W_{\nu}(\rho)=U_{\nu}(\rho)-\frac{\hbar^{2}}{2\mu}Q_{\nu\nu}(\rho)
\end{equation}
The diagonal second derivative non--adiabatic coupling term $Q_{\nu\nu}(\rho)$ is negative for every value of $\rho$, which makes the effective potential $W_{\nu}(\rho)$ more repulsive than $U_{\nu}(\rho)$. In treating the lowest channel in a single--channel calculation, a lower and upper bound to ground state energies or resonances can be determined by using either $U_{0}(\rho)$ or $W_{0}(\rho)$, respectively \cite{PhysRevA.19.1629}.

The fixed-$\rho$ adiabatic Hamiltonian in Eq. \eqref{eq:adiabatic_H} is diagonalized by expanding the channel function $\phi_{\nu}(\rho,\Omega)$ in a basis. The basis used in this work is the explicitly--correlated Gaussian (ECG) basis \cite{SV2,SV3}. In conjunction with treating the hyperradius as an adiabatic parameter, this basis is denoted the hyperradial explicitly--correlated Gaussian basis (CGHS). For an $N$--body system, the channels functions for a given total angular momentum and parity $J^{\pi}$ are expanded as follows,
\begin{subequations}
\label{eq:basis_expand}
\begin{multline}
\label{eq:basis_expand_simp}
\phi_{\nu}^{(LS)J^{\pi}T}(\rho,\Omega)\equiv\\\sum_{j}c_{j}^{(LS)J^{\pi}}e^{-\frac{1}{2}x^{T}A_{j}x}\left[\theta_{L}(u_{j,1},u_{j,2};x)\times\chi_{S}^{(Nb)}\right]_{JM_{J}}\chi_{TM_{T}}^{(Nb)}
\end{multline}
\begin{equation}
\label{eq:angl_func}
\theta_{LM_{L}}(u_{j,1},u_{j,2};x)=\left[Y_{L_{1}}(u_{1,j}^{T}x)\times Y_{L_{2}}(u_{2,j}^{T}x)\right]_{LM_{L}}
\end{equation}
\end{subequations}
where $\chi_{S}^{(Nb)}$ and $\chi_{TM_{T}}^{(Nb)}$ are $N$--body spin functions for the total spin and the total isospin, respectively.

\section{Interactions}
\label{sec:interactions}
Core--nucleon interactions have been used extensively in the few--body sector of nuclear physics, to treat heavier nuclei using few--body methods, which requires the development of realistic effective interactions between a core nucleus and a nucleon. Some interesting systems that are modeled by a core nucleus and few nucleons include some light nuclei and halo nuclei. 

\subsection{Two--body Interactions}
\label{subsec:core_nucleon_interaction}
\subsubsection{The \texorpdfstring{$\prescript{4}{}{\mathrm{He}}-n$}{text} Interaction}
\begin{figure}[H]
\centering
\subfigure[]{\includegraphics[width=0.48\columnwidth]{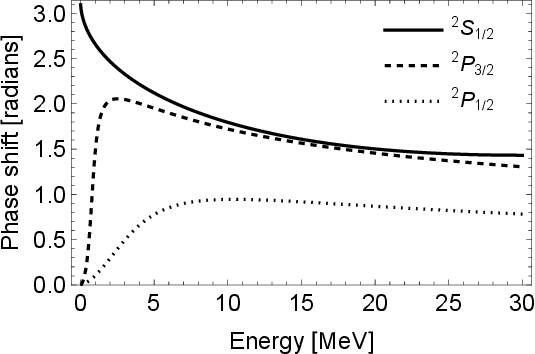}\label{fig:alpha_neutron_phaseshift_a}}
\subfigure[]{\includegraphics[width=0.48\columnwidth]{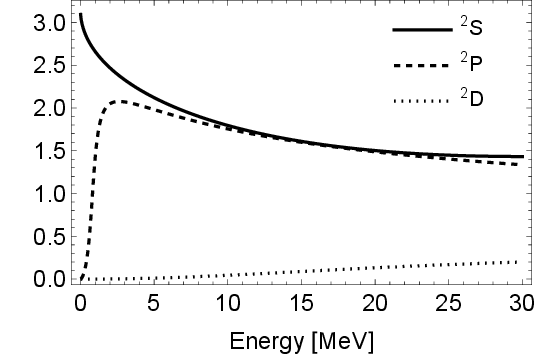}\label{fig:alpha_neutron_phaseshift_b}}
\caption{In (a) the $\prescript{4}{}{\mathrm{He}}-n$ phase shifts in the energy range of $0<E<30~\mathrm{MeV}$. The $s$--wave interaction is taken from \cite{PhysRevLett.79.2411} and the interactions for $l>0$, along with the spin--orbit interaction, are taken from \cite{PhysRevC.61.024318}. In (b) the $\prescript{4}{}{\mathrm{He}}-n$ phase shifts in the energy range of $0<E<30~\mathrm{MeV}$ for the simplified model, without a spin--orbit interaction. The $s$--wave and $d$--wave interactions are taken from \cite{PhysRevLett.79.2411} and the $p$--wave interaction is taken from \cite{PhysRevC.61.024318}.}
\label{fig:alpha_neutron_phaseshift}
\end{figure}

The $\prescript{4}{}{\mathrm{He}}-\mathrm{nucleon}$ effective interaction that is considered takes the form of Eq.~\eqref{eq:alpha_neutron_int_general}, which consists of a central interaction and a spin--orbit interaction \cite{GARRIDO1997153,10.1143/PTP.61.1327,PhysRevLett.79.2411,PhysRevC.61.024318}.
\begin{equation}
\label{eq:alpha_neutron_int_general}
V_{\alpha n}(\vec{r})=V_{\mathrm{c}}(r)+V_{\mathrm{s.o.}}(r)\left(\vec{l}\cdot\vec{s}\right)
\end{equation}
In the $\prescript{4}{}{\mathrm{He}}-n$ interaction given in Eq.~\eqref{eq:alpha_neutron_int_general}, the central and spin--orbit radial functions $V_{\mathrm{c}}(r)$ and $V_{\mathrm{s.o.}}(r)$ are represented either by a Woods--Saxon form \cite{PhysRevC.61.024318}, or by Gaussian's with parameters fitted to the experimental $\prescript{2}{}{S}_{1/2}$, $\prescript{2}{}{P}_{1/2}$ and $\prescript{2}{}{P}_{3/2}$ phase shifts in the energy range from $0<E<10~\mathrm{MeV}$. The main $\prescript{4}{}{\mathrm{He}}-n$ interaction used in this chapter is given in reference \cite{PhysRevC.61.024318}, where the Woods--Saxon radial function is fitted to a set of Gaussians. The corresponding phase shifts are shown in Fig. \ref{fig:alpha_neutron_phaseshift_a}.

The $s$--wave scattering length and effective range for the $\prescript{2}{}{S}_{1/2}$ state is $2.58~\mathrm{fm}$ and $1.38~\mathrm{fm}$, respectively \cite{GARRIDO1997153}. There is an $\prescript{4}{}{\mathrm{He}}-n$ resonance in the $\prescript{2}{}{P}_{3/2}$ state at an energy of $0.798~\mathrm{MeV}$ and a width of $0.648~\mathrm{MeV}$ \cite{TILLEY20023}. In order for an angular--momentum independent two--body interaction to reproduce the $\prescript{2}{}{S}_{1/2}$ scattering length and effective range as well as the $\prescript{2}{}{P}_{3/2}$ resonance position and width, there exists an artificial $\prescript{5}{}{\mathrm{He}}$ $s$--wave bound state that is deeply bound (with $E\approx-9.8~\mathrm{MeV}$ \cite{GARRIDO1997153,10.1143/PTP.61.1327,PhysRevLett.79.2411,PhysRevC.61.024318}). Our strategy to avoid having an artificial bound state is to adopt an angular momentum dependent semi--local central interaction, similar to what was implemented in \cite{GARRIDO1997153} and \cite{PhysRevLett.79.2411}. This simplifies the hyperspherical analysis of $\prescript{4}{}{\mathrm{He}}-\mathrm{nucleon}$ systems, because it eliminates the nonphysical hyperspherical channels from the computation. The modified central interaction adopted has the form given in Eq.~\eqref{eq:angl_mom_int},
\begin{equation}
\label{eq:angl_mom_int}
V_{\mathrm{c}}(\vec{r})=\sum_{l,m}V_{c}^{(l)}(r)\vert lm\rangle\langle lm\vert
\end{equation}
where the $s$--wave component ($l$=0) is a repulsive interaction to reproduce the $\prescript{2}{}{S}_{1/2}$ phase shift, thus eliminating any $s$--wave bound state from appearing. This angular momentum component is chosen to be the $s$--wave component in \cite{PhysRevLett.79.2411}. For angular momentum states with $l>0$, the interaction is chosen to be central interaction in \cite{PhysRevC.61.024318} that reproduces the $\prescript{2}{}{P}_{1/2}$ and $\prescript{2}{}{P}_{3/2}$ phase shifts along with the $\prescript{5}{}{\mathrm{He}}$ resonance and width.

Another form of the $\prescript{4}{}{\mathrm{He}}-\mathrm{nucleon}$ interaction considered is a simplified version of the one in Eq. \eqref{eq:alpha_neutron_int_general}. Namely, it takes the following form,
\begin{equation}
\label{eq:alpha_neutron_simple}
V_{\alpha n}(\vec{r})=V_{\mathrm{c}}(r)
\end{equation}
where there is no spin--orbit interaction and the central interaction is angular momentum--dependent that includes up to $l=2$. The remaining partial waves take the form of the interaction set to give $l=2$ phase shifts. The form and parameters used in the $\prescript{4}{}{\mathrm{He}}-n$ interaction are given in Table \ref{table:two_body_table_alpha_neutron}. The $\prescript{4}{}{\mathrm{He}}+n$ phase shifts for up to $l=2$ using this interaction are shown in Figure \ref{fig:alpha_neutron_phaseshift_b}.
\begin{table}[H]
\caption{Parameters for the $\prescript{4}{}{\mathrm{He}}-n$ interaction. The $\prescript{4}{}{\mathrm{He}}-n$ parameters are taken from \cite{PhysRevLett.79.2411,PhysRevC.61.024318}.}
\label{table:two_body_table_alpha_neutron}
\begin{ruledtabular}
\begin{tabular}{ccc}
    \multicolumn{3}{c}{\bf{$\prescript{4}{}{\mathrm{He}}-n$ Interaction}}\\
    \hline
    $\mathrm{Partial~Wave}~(l)$ & $V_{0}^{(l)}~[\mathrm{MeV}]$ & $r_{0}^{(l)}~[\mathrm{fm}]$\\
    \hline
    $l=0$ & $48.2$ & $2.33$\\
    $l=1$ & $-51.675$ & $2.33$\\
    $l\geq2$ & $-21.93$ & $2.03$\\
\end{tabular}
\end{ruledtabular}
\end{table}
In this simplified two--body model for the $\prescript{4}{}{\mathrm{He}}-n$ interaction, the $j=1/2$ and $j=3/2$ states are degenerate due to the absence of a spin--orbit interaction term. This leads to a $j=1/2$ interaction that is more attractive than in more realistic models, where the $j=1/2$ potentials are more repulsive due to the spin--orbit interaction. This extra attraction in the $j=1/2$ potentials are approximately compensated, in systems with more than one neutron, by introducing a spin--dependent $\prescript{4}{}{\mathrm{He}}-nn$ three--body interaction, as is discussed in the next section.
\subsubsection{Neutron--neutron Interaction}
The $nn$ interaction used is a spin--dependent central interaction whose general form is given in Eq.~\eqref{eq:nn_gauss_int}.
\begin{multline}
\label{eq:nn_gauss_int}
V_{nn}(\vec{r}_{ij})=V_{0}^{nn}(\vec{r}_{ij})|\chi_{0}^{(ij)}\rangle\langle\chi_{0}^{(ij)}|\\+V_{1}^{nn}(\vec{r}_{ij})|\chi_{1}^{(ij)}\rangle\langle\chi_{1}^{(ij)}|
\end{multline}
In Eq. \eqref{eq:nn_gauss_int}, $V_{0}^{nn}(\vec{r})$ and $V_{1}^{nn}(\vec{r})$ are the singlet and triplet potentials respectively. The spinors $\chi_{0}^{(ij)}$ and $\chi_{1}^{(ij)}$ are the two--body $nn$ spin functions for total spin singlet and triplet states, respectively. The form of the radial functions in Eq. \eqref{eq:nn_gauss_int} are Gaussian functions for both the singlet and triplet potentials, whose parameters are fit to reproduce low energy scattering phase shifts. The Gaussian parameters are provided in Table~\ref{table:two_body_table}.
\begin{table}[H]
\caption{Parameters for the $nn$ interaction. The $nn$ parameters are taken from \cite{PhysRevLett.125.052501,PhysRevC.103.024004}.}
\label{table:two_body_table}
\begin{ruledtabular}
\begin{tabular}{ccc}
    \multicolumn{3}{c}{\bf{$nn$ Interaction}}\\
    \hline
    $\mathrm{Spin~State}~(s)$ & $V_{0}^{(s)}~[\mathrm{MeV}]$ & $r_{0}^{(s)}~[\mathrm{fm}]$\\
    \hline
    $s=0$ & $-31.7674$ & $1.7810$\\
    $s=1$ & $95.7280$ & $0.8809$\\
\end{tabular}
\end{ruledtabular}
\end{table}
This simple model for the $nn$ interaction has proven to reproduce quite well the low energy physics for the $3n$ and $4n$ systems \cite{PhysRevLett.125.052501,PhysRevC.103.024004}. Owing to the large $s$--wave scattering length ($a_{s}^{nn}\approx-18~\mathrm{fm}$), systems of interacting neutrons exhibit universal physics that is largely insensitive of the details of the two--body interaction. This $nn$ interaction has been shown to give results that are in quantitative agreement with more realistic phenomenological nuclear models like the Argonne V potentials \cite{PhysRevC.51.38}. Thus, for this work we use this simplified Gaussian form of the $nn$ interaction.
\subsection{Three--body Interactions}
\begin{figure}[ht!]
\centering
%\subfigure[]{\includegraphics[width=4.25 cm]{three_body_singlet.eps}}\label{fig:he6_v3_a}
%\subfigure[]{\includegraphics[width=4.25 cm]{three_body_triplet.eps}}\label{fig:he6_v3_b}
\includegraphics[width=8.6 cm]{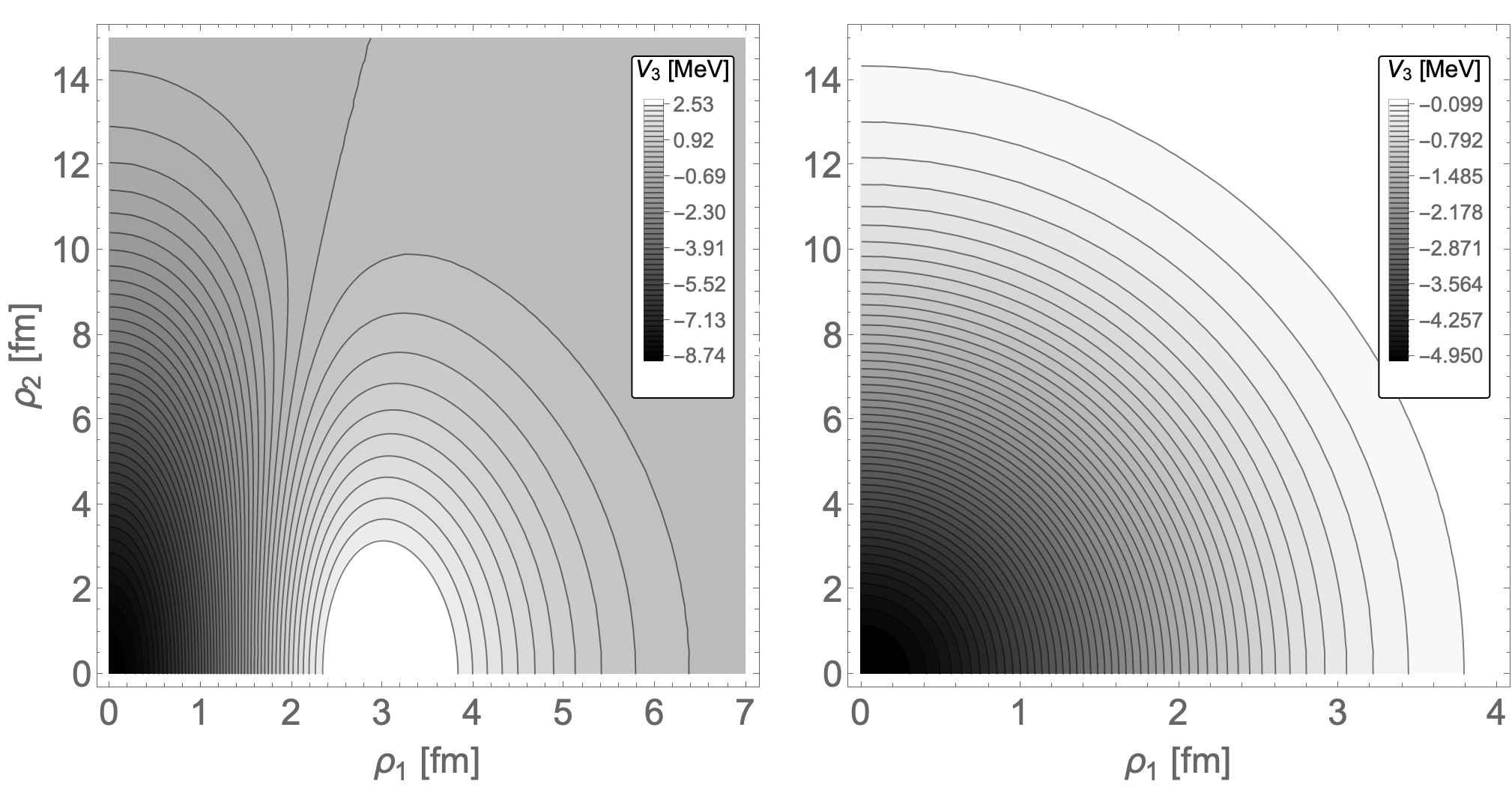}
\caption{\label{fig:helium6_three_body_plots}The spin singlet (left panel) and triplet (right panel) three--body interaction potentials for the $\prescript{4}{}{\mathrm{He}}-nn$ subsystem. These potentials are plotted as a function of the magnitude of three--body Jacobi vectors $\vec{\rho}_{1}$ and $\vec{\rho}_{2}$. The magnitude of $\vec{\rho}_{1}$ is proportional to the $nn$ inter-particle distance and $\vec{\rho}_{2}$ is the vector connecting the center of mass of the $nn$ subsystem to the $\prescript{4}{}{\mathrm{He}}$ nucleus. The interaction parameters for the singlet interaction were tuned to reproduce the $\prescript{6}{}{\mathrm{He}}(0^+)$ ground state energy and the $\prescript{6}{}{\mathrm{He}}(2^+)$ resonance. The parameters for the triplet interaction were tuned to reproduce the $\prescript{8}{}{\mathrm{He}}(0^+)$ ground state energy and rms matter radius.}
\end{figure}

For more than one neutron interacting with an $\prescript{4}{}{\mathrm{He}}$ nucleus, there should be a three--body force introduced to capture the complicated interaction between the $\prescript{4}{}{\mathrm{He}}$ nucleus and two neutrons. The interactions treated in the $\prescript{4}{}{\mathrm{He}}-nn$ system are the two--body $nn$ interaction, the two--body $\prescript{4}{}{\mathrm{He}}-n$ interaction, and a three--body interaction. The full inter-particle interaction in the three--body Hamiltonian is given in Eq.~\eqref{eq:he6_interaction_H} as,
\begin{equation}
\label{eq:he6_interaction_H}
\hat{V}_{\mathrm{int.}}=V_{nn}(\vec{r}_{12})+V_{\alpha n}(\vec{r}_{13})+V_{\alpha n}(\vec{r}_{23})+V_{3b}(\vec{r}_{12},\vec{r}_{13},\vec{r}_{23}).
\end{equation}
In Eq. \eqref{eq:he6_interaction_H} the subscript $\alpha$ is used as a label for the $\prescript{4}{}{\mathrm{He}}$ nucleus. The $\prescript{4}{}{\mathrm{He}}-n$ interaction used is described in the previous section and given by Eqs. \eqref{eq:angl_mom_int} and \eqref{eq:alpha_neutron_simple}. The three--body interaction $V_{3b}$ is a spin--dependent interaction that depends on the composite spin of the $\prescript{4}{}{\mathrm{He}}-nn$ system and given in Eq.~\eqref{eq:gauss_3B_int} as
\begin{multline}
\label{eq:gauss_3B_int}
V_{3b}(\vec{r}_{ij},\vec{r}_{ik},\vec{r}_{jk})=V_{0}(\vec{r}_{ij},\vec{r}_{ik},\vec{r}_{kj})|\chi_{0}^{(ijk)}\rangle\langle\chi_{0}^{(ijk)}|\\+V_{1}(\vec{r}_{ij},\vec{r}_{ik},\vec{r}_{kj})|\chi_{1}^{(ijk)}\rangle\langle\chi_{1}^{(ijk)}|
\end{multline}
where $\chi_{0}^{(ijk)}$ and $\chi_{1}^{(ijk)}$ are the three--body spinors for the total spin singlet and triplet states, respectively. The spacial dependence for the spin--singlet and triplet components are represented by a set of Gaussian functions of the following form,
\begin{equation}
\label{eq:gauss_3B_form}
V_{S}(\vec{r}_{ij},\vec{r}_{ik},\vec{r}_{jk})=\sum_{m=1}^{n}V_{m}^{(S)}e^{-\alpha_{m}^{(S)}r_{ij}^{2}-\beta_{m}^{(S)}(r_{ik}^{2}+r_{jk}^{2})}.
\end{equation}
The parameters for the singlet and triplet three--body interactions are given in Table \ref{table:three_body_table} and plots of the potentials are shown in Figure \ref{fig:helium6_three_body_plots}.
\begin{table}[ht!]
\caption{Parameters for the spin--dependent three--body interaction between the $\prescript{4}{}{\mathrm{He}}-nn$ triad. Note that ${r_\alpha}_{m}^{(S)}=1/\sqrt{\alpha_{m}^{S}}$ and ${r_\beta}_{m}^{(S)}=1/\sqrt{\beta_{m}^{S}}$.}
\label{table:three_body_table}
\begin{ruledtabular}
\begin{tabular}{cccc}
    \multicolumn{4}{c}{\bf{Singlet Interaction}}\\
    \hline
    $m$ & $V_{m}^{(0)}~[\mathrm{MeV}]$ & ${r_\alpha}_{m}^{(0)}~[\mathrm{fm}]$ & ${r_\beta}_{m}^{(0)}~[\mathrm{fm}]$\\
    \hline
    $m=1$ & $-20.8792$ & $2.47917$ & $7.66667$\\
    $m=2$ & $12.0000$ & $4.50000$ & $7.00000$\\
    \hline
    \multicolumn{4}{c}{\bf{Triplet Interaction}}\\
    \hline
    $m$ & $V_{m}^{(1)}~[\mathrm{MeV}]$ & ${r_{\alpha}}_{m}^{(1)}~[\mathrm{fm}]$ & ${r_{\beta}}_{m}^{(1)}~[\mathrm{fm}]$\\
    \hline
    $m=1$ & $-5.08000$ & $2.50000$ & $8.00000$\\
\end{tabular}
\end{ruledtabular}
\end{table}

The Gaussian parameters in Table \ref{table:three_body_table} were chosen to reproduce properties for both $\prescript{6}{}{\mathrm{He}}$ and $\prescript{8}{}{\mathrm{He}}$ nuclei. The spin--singlet three--body parameters were chosen specifically to reproduce the $\prescript{6}{}{\mathrm{He}}(0^{+})$ ground--state energy of $-0.973~\mathrm{MeV}$ and the $\prescript{6}{}{\mathrm{He}}(2^{+})$ resonance and width. The spin--triplet three--body parameters were chosen to reproduce the $\prescript{8}{}{\mathrm{He}}(0^{+})$ ground--state energy of $-3.11~\mathrm{MeV}$ \cite{TILLEY20023,TILLEY2004155,PhysRev.139.B592}, along with other $\prescript{8}{}{\mathrm{He}}$ properties like the rms matter radius and the rms $\prescript{4}{}{\mathrm{He}}-4n$ separation \cite{TANIHATA1988592, WAKASA2022105329}. With the two--body and three--body interactions defined for the $\prescript{4}{}{\mathrm{He}}-n$, $nn$ and $\prescript{4}{}{\mathrm{He}}-nn$ systems, interacting systems consisting of an $\prescript{4}{}{\mathrm{He}}$ nucleus and a few neutrons are investigated using the adiabatic hyperspherical representation. The Hamiltonian under investigation contain only spin--dependent central interactions, with no other interactions that couple spin and angular momentum, so the total orbital angular momentum $L$ and the total spin $S$ are good quantum numbers. In the following sections, structure properties and collision properties are investigated for helium isotopes with $N>2$ in different $(L^{\pi},S)J^{\pi}$ symmetry states. 

\section{Results}
\label{sec:results}
\subsection{Bound and Resonant States of the \texorpdfstring{$\prescript{6}{}{\mathrm{He}}$}{text} Nucleus}
\label{sec:three_body_sys}
\begin{figure}[ht!]
% Center the entire figure (containing the two subfigures).
\centering
\subfigure[]{\includegraphics[width=8.6 cm]{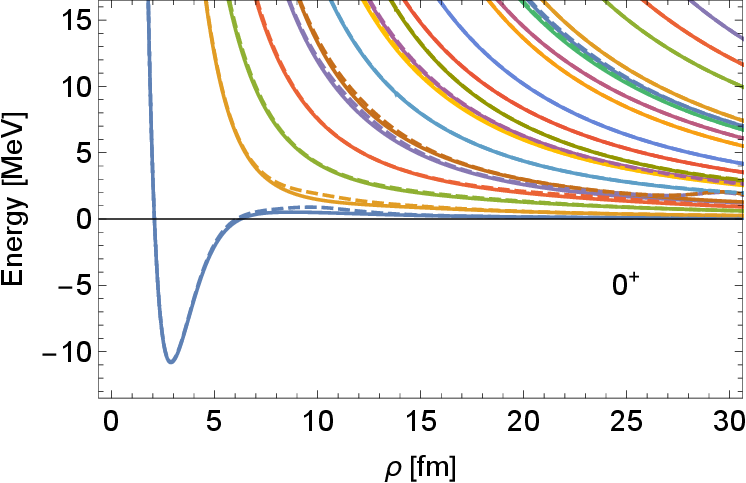}\label{fig:he6_j0_pots}}

\subfigure[]{\includegraphics[width=8.6 cm]{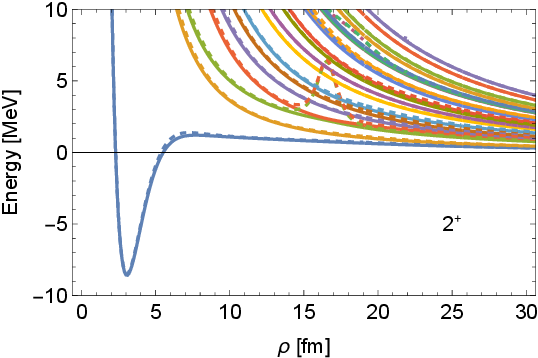}\label{fig:he6_j2_pots}}
\caption{\label{fig:helium6_J0_J2}The lowest few adiabatic hyperspherical potential curves for the $\prescript{4}{}{\mathrm{He}}-nn$ system are displayed in the $(L^{\pi},S)J^{\pi}=(0^{+},0)0^{+}$ symmetry in (a) and $(L^{\pi},S)J^{\pi}=(2^{+},0)2^{+}$ in (b). The solid curves are the adiabatic potentials $U_{\nu}(\rho)$ and the dashed curves include the second--derivative non--adiabatic coupling term. The visible peaks in the second and third potentials in (b) are due to the avoided crossing clearly seen near $\rho=17~\mathrm{fm}$.}
\end{figure}

The lowest few hyperradial potential curves for the $\prescript{4}{}{\mathrm{He}}-nn$ system in the $(L^{\pi},S)J^{\pi}=(0^{+},0)0^{+}$ and $(2^{+},0)2^{+}$ symmetries are shown in Fig.~\ref{fig:helium6_J0_J2}. In Fig.~\ref{fig:he6_j0_pots}, the $0^{+}$ hyperradial potentials are shown with and without the diagonal second--derivative coupling matrix element included. Likewise, in Fig.~\ref{fig:he6_j2_pots}, the $2^{+}$ hyperradial potentials are shown with and without the diagonal correction term included. In both figures, the effective hyperradial potentials (those with the diagonal correction term) are represented as dashed lines and shown for the lowest six hyperradial channels. For the $0^{+}$ symmetry, the lowest hyperradial potential curve in Fig. \ref{fig:he6_j0_pots} has a potential well with a minimum below the three--body breakup threshold at $E=0$, which supports a bound state of the $\prescript{4}{}{\mathrm{He}}-nn$ system in this symmetry. This bound--state is the $\prescript{6}{}{\mathrm{He}}(0^{+})$ ground--state with a corresponding energy of $-0.98~\mathrm{MeV}$, based on the fit of the spin--singlet three--body term in Eq.~\eqref{eq:gauss_3B_int} and a full diagonalization calculation using 100 basis functions. The bound--state calculation was confirmed by also performing a hyperradial coupled--channel calculation including up to 8 channels and integrating over the range of $0.1~\mathrm{fm}\leq\rho\leq70~\mathrm{fm}$. The results for the bound--state energy for different numbers of included channels are presented in the second and third columns of Table \ref{table:three_body_J0_table}.
\begin{table}[hb!]
\caption{The $\prescript{6}{}{\mathrm{He}}$ $0^{+}$ ground--state energy $E_{\mathrm{g.s.}}$, and the rms matter radius\footnote{The rms matter radius is computed as follows, $\langle r^2\rangle^{1/2}_{\mathrm{rms}}=\left[\frac{\mu}{M_{\mathrm{T}}}\langle\rho^2\rangle+\frac{M_{A}}{M_{\mathrm{T}}}\langle r^2_{A}\rangle_{\mathrm{rms}}\right]^{1/2}$, where $M_{T}$ is the total mass of the system, $\langle\rho^{2}\rangle$ is the expectation value of the square of the hyperradius and $\langle r^2_{A}\rangle$ is the square of the rms radius of subsystem $A$ with mass $M_{A}$.} $\langle r^{2}\rangle^{\frac{1}{2}}_{\mathrm{rms}}$ are shown in the second and third columns. The $2^{+}$ resonance energy $E_{\mathrm{res.}}$, and width $\Gamma_{\mathrm{res.}}$ are given in the last two columns. Each row is a calculation solving Eq. \eqref{eq:coupled_channel_eqs} for a set number of included channels, with the number denoted in the first column. The computed rms matter radius agrees to within the error bounds of the measured value of $2.48(3)~\mathrm{fm}$, extracted from \cite{TANIHATA1988592}. The rms matter radius for the $\prescript{4}{}{\mathrm{He}}$ nucleus is taken to be 1.57~fm \cite{TANIHATA1992261,WAKASA2022105329}.}
\label{table:three_body_J0_table}
\begin{ruledtabular}
\begin{tabular}{ccc|cc}
     & \multicolumn{2}{c}{\bf{$\prescript{6}{}{\mathrm{He}}(0^{+})$}} & \multicolumn{2}{c}{\bf{$\prescript{6}{}{\mathrm{He}}(2^{+})$}} \\
    \hline
    $\mathrm{No.}$ & $E_{\mathrm{g.s.}}~[\mathrm{MeV}]$ & $\langle r^{2}\rangle^{\frac{1}{2}}_\mathrm{rms}~[\mathrm{fm}]$ & $E_{\mathrm{res.}}~[\mathrm{MeV}]$ & $\Gamma_{\mathrm{res.}}~[\mathrm{MeV}]$\\
    \hline
    $1$ & $-0.8063$ & $2.5497$ & $0.8423$ & $0.1036$\\
    $2$ & $-0.9586$ & $2.5132$ & $0.8225$ & $0.1068$\\
    $3$ & $-0.9655$ & $2.5121$ & $0.8162$ & $0.1014$\\
    $4$ & $-0.9751$ & $2.5086$ & $0.8110$ & $0.1008$\\
    $5$ & $-0.9777$ & $2.5075$ & $0.8039$ & $0.0973$\\
    $6$ & $-0.9793$ & $2.5072$ & $0.8019$ & $0.0957$\\
    $7$ & $-0.9798$ & $2.5070$ & $0.8013$ & $0.0948$\\
    $8$ & $-0.9799$ & $2.5070$ & $0.8010$ & $0.0926$\\
    %$9$ & $-0.9801$ & $2.5069$\\
    %$10$ & $-0.9801$ & $2.5069$\\
\end{tabular}
\end{ruledtabular}
\end{table}

\begin{figure}[ht!]
% Center the entire figure (containing the two subfigures).
\centering
\includegraphics[width=8.6 cm]{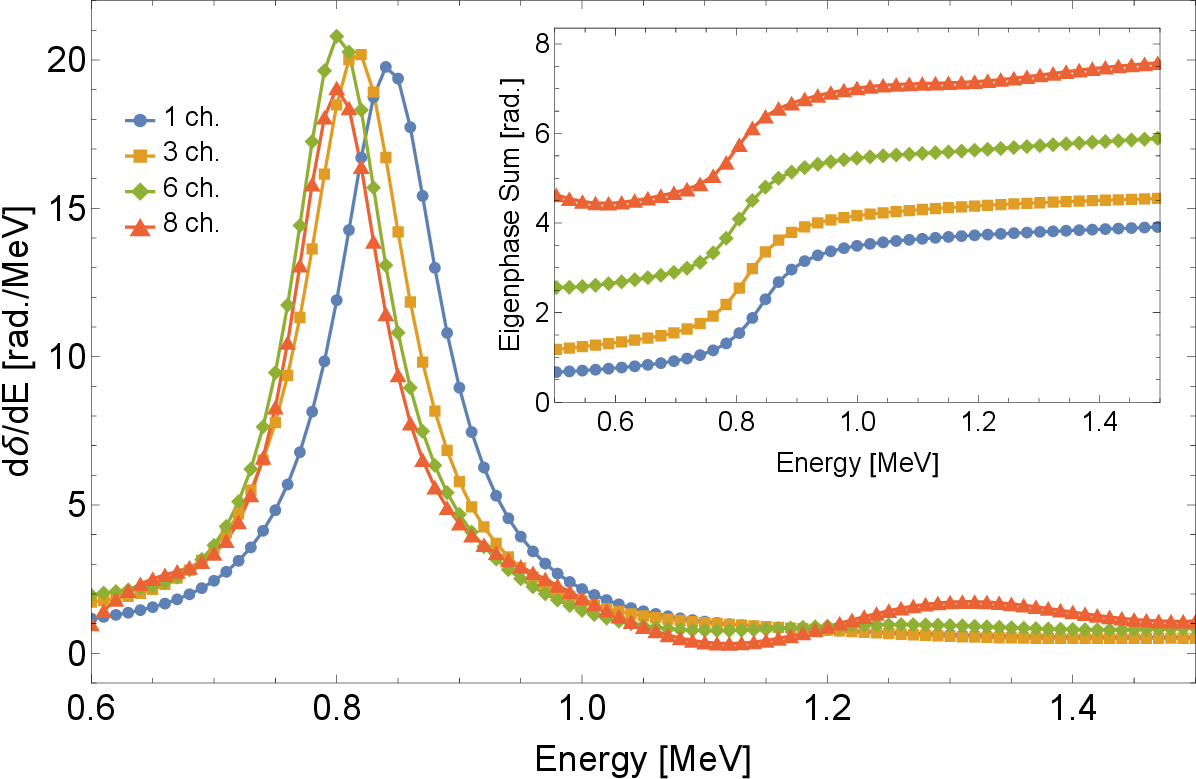}
%\subfigure[]{\includegraphics[width=8.6 cm]{he6_eigenphase_deriv_8ch_fortran.eps}\label{fig:he6_eph_b}}
\caption{\label{fig:helium6_eigenphase_plots}The sum of the eigenphase shifts (see inset) and their energy derivatives (see main figure) are shown for the $\prescript{6}{}{\mathrm{He}}(2^{+})$ system over an energy range from $0.6~\mathrm{MeV}$ to $1.4~\mathrm{MeV}$. Each curve represents a calculation performed by solving Eq. \eqref{eq:coupled_channel_eqs} where 1, 3, 6, and 8 channels are included. There is a clear peak in the energy derivative at $0.80~\mathrm{MeV}$ with a corresponding increase in phase of $\approx2$ radians, indicating the presence of the $2^{+}$ resonance state of $\prescript{6}{}{\mathrm{He}}$.}
\end{figure}

In Fig.~\ref{fig:he6_j2_pots}, the lowest $2^{+}$ hyperradial potential curve has features that indicate the support for a three--body shape--resonance above the $\prescript{4}{}{\mathrm{He}}+n+n$ continuum. This hyperradial potential curve has a local minimum at $\rho\approx2.8~\mathrm{fm}$ and a barrier with a height of $\approx2~\mathrm{MeV}$, with its maximum located at $\rho\approx~6.5~\mathrm{fm}-7~\mathrm{fm}$. The barrier peak is formed due to the large amount of attraction from the $p$--wave component $\prescript{4}{}{\mathrm{He}}-n$ interaction, primarily contributing from the $\prescript{2}{}{P}_{3/2}$ state, in conjunction with the hyperradial momentum barrier that goes as $U(\rho)\rightarrow\frac{\hbar^{2}3.75}{2\mu\rho^{2}}$ at large $\rho$. The other hyperradial potential curves describe hyperradial channels that at large $\rho$, describe various three--body $\prescript{4}{}{\mathrm{He}}+n+n$ continuum states. The $2^{+}$ resonance is computed by using the slow--variable discretization (SVD) method \cite{Tolstikhin1996} in conjunction with the discrete variable representation (DVR) to propagate the $R$--matrix \cite{Wang,MANOLOPOULOS198823}. The results for a single--channel calculation up to an 8--channel calculation, propagating the $R$--matrix from $0.1~\mathrm{fm}\leq\rho\leq100~\mathrm{fm}$ using 1000 DVR sectors are shown in Fig.~\ref{fig:helium6_eigenphase_plots}.

In the inset of Fig.~\ref{fig:helium6_eigenphase_plots}, the sum of the eigenphase shifts for calculations with up to 8 coupled channels are shown for an energy range $0.6~\mathrm{MeV}\leq E\leq1.5~\mathrm{MeV}$. The eigenphase sum shows a clear resonance feature in the energy region from $0.7~\mathrm{MeV}\leq E\leq 0.9~\mathrm{MeV}$. In this region, the eigenphase sum rises by approximately 2 radians. In addition, there is a clear peak in this region when looking at the energy--derivative shown in the main plot of Fig \ref{fig:helium6_eigenphase_plots}. This peak is fitted to a Lorentzian, with key parameters being the resonance position $E_{\mathrm{res.}}$, and the width of the resonance $\Gamma_{\mathrm{res.}}$. The energy and width of the $2^{+}$ state computed from solving the coupled hyperradial Schr$\ddot{\mathrm{o}}$dinger equation for up to 8 included channels are given in the last two columns of Table \ref{table:three_body_J0_table}. From this fitting procedure, the resonance and width of the $2^{+}$ state from the 8--channel calculation are $E_{\mathrm{res.}}=0.8010~\mathrm{MeV}$ and $\Gamma=0.0926~\mathrm{MeV}$. The experimental values for the $2^{+}$ resonance and width are $E_{\mathrm{res.}}=0.822(25)~\mathrm{MeV}$ and $\Gamma_{\mathrm{res.}}=0.113(20)~\mathrm{MeV}$ \cite{TILLEY20023}. The calculated results are in agreement with experiment to within experimental error.

\begin{figure}[ht!]
% Center the entire figure (containing the two subfigures).
\centering
\includegraphics[width=8.6 cm]{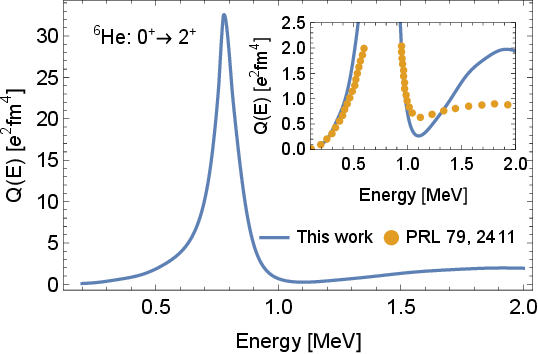}
\caption{\label{fig:he6_quad_transition} The square of the quadrupole transition between the $\prescript{6}{}{\mathrm{He}}$ $0^{+}$ ground state and the $\prescript{6}{}{\mathrm{He}}$ $2^{+}$ continuum states at some final state energy $E$. The peak in the transition matrix element corresponds to the resonance position of the $2^{+}$ state and the FWHM is on the order of the width of the resonance. The inset shows a comparison with a previous hyperspherical calculation \cite{PhysRevLett.79.2411}.}
\end{figure}

With the $\prescript{6}{}{\mathrm{He}}$ $0^{+}$ ground state wave function and the $2^{+}$ wave function at a given scattering energy $E$, transition matrix elements between the two states can be computed. The $0^{+}$ state of $\prescript{6}{}{\mathrm{He}}$ can undergo a transition to the $2^{+}$ state through an electric quadrupole transition. The electric quadrupole operator is given in Eq. \eqref{eq:quad_operator} as
\begin{equation}
\label{eq:quad_operator}
\hat{Q}_{2m}=\sum_{i=1}^{N}q_{i}r_{i}^{2}Y_{2m}(\hat{r}_{i})
\end{equation}
where $q_{i}$ is the charge and $\vec{r}_{i}$ is the position vector of the $i^{\mathrm{th}}$ particle. For the $\prescript{6}{}{\mathrm{He}}$ system, the quadrupole operator reduces to $\hat{Q}_{2m}=2er_{\alpha}^{2}Y_{2m}(\hat{r}_{\alpha})$ since the neutron has no charge and the charge of the $\prescript{4}{}{\mathrm{He}}$ nucleus is $q_{\alpha}=2e$. In Figure \ref{fig:he6_quad_transition}, the absolute square of the electric quadrupole transition matrix element, $Q(E)=|\langle0^{+}|\hat{Q}|2^{+}\rangle|^{2}$, is plotted as a function of energy.
The energy dependent quadrupole transition matrix element exhibits a maximum value of 32.6 $e^{2}\mathrm{fm}^{4}$, as shown in Fig. \ref{fig:he6_quad_transition}. The peak in the quadrupole transition matrix element is due to the resonance of the $2^{+}$ state and is located at an energy close to the resonance energy to within the resonance width of $0.1~\mathrm{MeV}$. The inset displays results from a previous hyperspherical calculation \cite{PhysRevLett.79.2411} on top of the results from this work over a smaller range in the quadrupole transition matrix element. The comparison shows qualitative agreement amongst the two calculations with reasonable quantitative agreement near the peak. Differences between the two calculations can be attributed to the slightly different nuclear interactions used and also on different levels of convergence in the computed hyperspherical eigenvalues and eigenvectors.

\subsection{Resonant and Scattering States of the \texorpdfstring{$\prescript{7}{}{\mathrm{He}}$ Nucleus}{text}}
\label{sec:four_body_sys}
In the previous section, the lowest bound and resonant states of the $\prescript{6}{}{\mathrm{He}}$ system were computed with the nuclear Hamiltonian given in Eq. \eqref{eq:he6_interaction_H}. The method can naturally be extended to the next isotope, the $\prescript{7}{}{\mathrm{He}}$ nucleus, modeled as $\prescript{4}{}{\mathrm{He}}+3n$. The $\prescript{7}{}{\mathrm{He}}$ nucleus is unbound but has a ${3/2}^{-}$ resonance in the $\prescript{6}{}{\mathrm{He}}+n$ two--body continuum. The experimental resonance and width of this state is $E_{\mathrm{res.}}=0.446$~MeV and $\Gamma_{\mathrm{res.}}=0.183$~MeV \cite{BECK2007128}.

\begin{figure}[ht!]
% Center the entire figure (containing the two subfigures).
\centering
\subfigure[]{\includegraphics[width=8.6 cm]{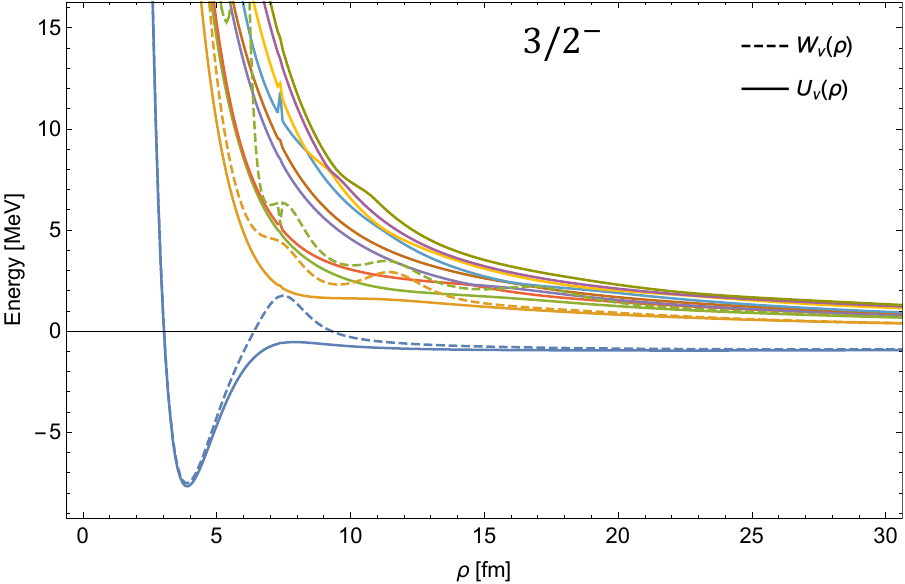}\label{fig:he7_pot_a}}
\subfigure[]{\includegraphics[width=8.6 cm]{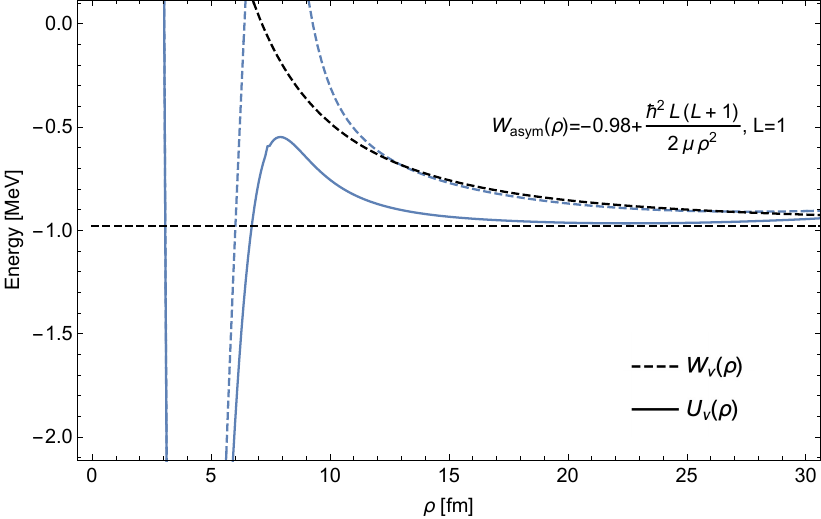}\label{fig:he7_pot_b}}
\caption{\label{fig:helium7_potentials}The lowest few adiabatic potential curves for the $\prescript{4}{}{\mathrm{He}}+3n$ system in the $(L,S)J^{\pi}=(1^{-},1/2){3/2}^{-}$ symmetry are shown in (a). In (b), a the lowest hyperradial potential is shown on a zoomed--in scale, highlighting the asymptotic form of this potential at large hyperradius. The solid curves are the adiabatic potentials, and the dashed curves are the effective hyperradial potentials, which include the second--derivative non--adiabatic coupling term.}
\end{figure}

The $\prescript{4}{}{\mathrm{He}}+3n$ system in the $(1^{-},1/2)3/2^{-}$ symmetry is treated in the adiabatic hyperspherical approach. The lowest 10 hyperradial potential curves for this system are shown in Figure \ref{fig:helium7_potentials}. The lowest potential curve goes to the $\prescript{6}{2}{\mathrm{He}(0^{+})}+n$ threshold at large $\rho$, where the system fragments through a $p$--wave interaction between the $\prescript{6}{}{\mathrm{He}}$ nucleus and the free neutron. This behavior is highlighted in Fig. \ref{fig:helium7_potentials}(b). In Fig. \ref{fig:helium7_potentials}(b), the lowest hyperradial potential curve with (solid) and without (dashed) the second derivative diagonal coupling term is shown on an energy scale that highlights the asymptotic behavior (black--dashed) at large hyperradius. The lowest hyperradial potential matches the expected asymptotic behavior for $\rho>12~\mathrm{fm}$.

\begin{figure}[ht!]
\centering
\includegraphics[width=0.9\columnwidth]{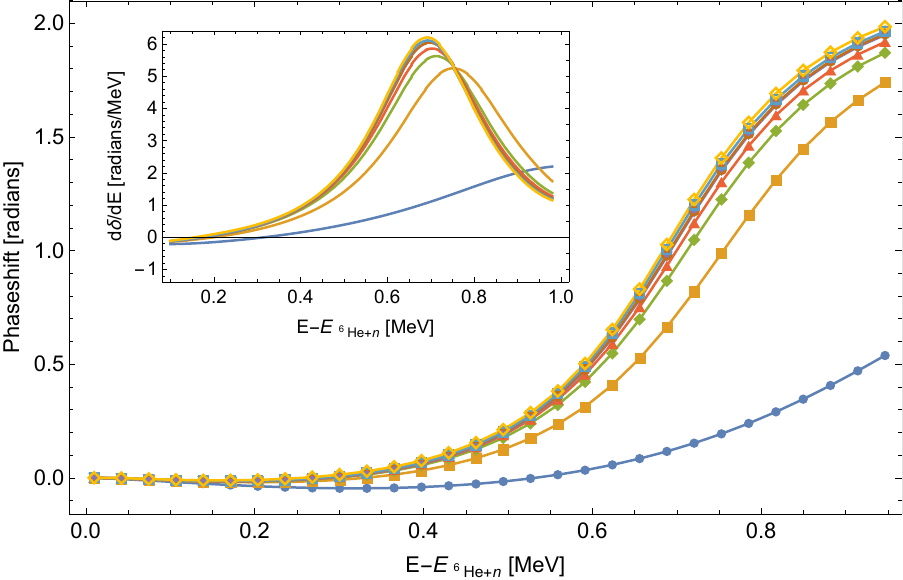}
\caption{The $\prescript{6}{}{\mathrm{He}}(0^{+})+n$ phase shift for the $3/2^{-}$ symmetry in the energy range of $0<E<0.97~\mathrm{MeV}$ above the two--body continuum threshold for the simplified $\prescript{4}{}{\mathrm{He}}-n$ interaction model. Each curve represents a different number of included channels when solving Eq. \eqref{eq:coupled_channel_eqs}. The inset shows the energy derivative of the phase shift, which peaks at the resonance energy.}
\label{fig:he7_phaseshift_simp}
\end{figure}

To compute the $\prescript{7}{}{\mathrm{He}}(3/2^{-})$ resonance, the coupled hyperradial 
Shr$\mathrm{\ddot{o}}$dinger equation, Eq. \eqref{eq:coupled_channel_eqs}, is solved numerically. The energy range for which the $\prescript{6}{}{\mathrm{He}}(0^{+})+n$ phase shift is computed is $0.01~\mathrm{MeV}<E<0.97~\mathrm{MeV}$ above the two--body continuum threshold. The phase shift and its energy derivative are shown in Figure \ref{fig:he7_phaseshift_simp} for up to 8 included channels. As the number of included channels increases, the phase shift rises more rapidly at a lower and lower energy, indicating the position of the resonance shifts to the left and the width of the resonance narrows. The position and width are extracted from the data by fitting the energy derivative to a Lorentzian near the peak. Table \ref{table:three_body_phase_table_he7} gives the resonance position and width of the $\prescript{7}{}{\mathrm{He}}(3/2^{-})$ state for different number of included channels. In the model Hamiltonian used in this work, the $\prescript{7}{}{\mathrm{He}}(3/2^{-})$ resonance and width is calculated by extrapolating to an infinite number of coupled channels, which gives values of $E_{\mathrm{res.}}=0.692~\mathrm{MeV}$ and $\Gamma_{\mathrm{res.}}=0.304~\mathrm{MeV}$, respectively. The resonance and width are within $55\%-60\%$ of the experimental values. 

\begin{table}[ht!]
\caption{The $\prescript{7}{}{\mathrm{He}}$ ${3/2}^{-}$ resonance energy $E_{\mathrm{res.}}$ and width $\Gamma_{\mathrm{res.}}$ are tabulated for a given number of included channels in solving the coupled hyperradial Shr$\ddot{\mathrm{o}}$dinger equation. The energies and widths were extracted from fitting the energy derivative of the eigenphase sum to a Lorentzian.}
\label{table:three_body_phase_table_he7}
\begin{ruledtabular}
\begin{tabular}{ccc}
    $\mathrm{No.~of~Channels}$ & $E_{\mathrm{res.}}~[\mathrm{MeV}]$ & $\Gamma_{\mathrm{res.}}~[\mathrm{MeV}]$\\
    \hline
    $1$ & $1.010402$ & $0.798171$\\
    $2$ & $0.752725$ & $0.359007$\\
    $3$ & $0.713907$ & $0.338151$\\
    $4$ & $0.703771$ & $0.325007$\\
    $5$ & $0.697796$ & $0.314359$\\
    $6$ & $0.697291$ & $0.314350$\\
    $7$ & $0.693488$ & $0.310499$\\
    $8$ & $0.692128$ & $0.304289$\\
\end{tabular}
\end{ruledtabular}
\end{table}

\begin{figure}[ht!]
% Center the entire figure (containing the two subfigures).
\centering
\includegraphics[width=1.0\columnwidth]{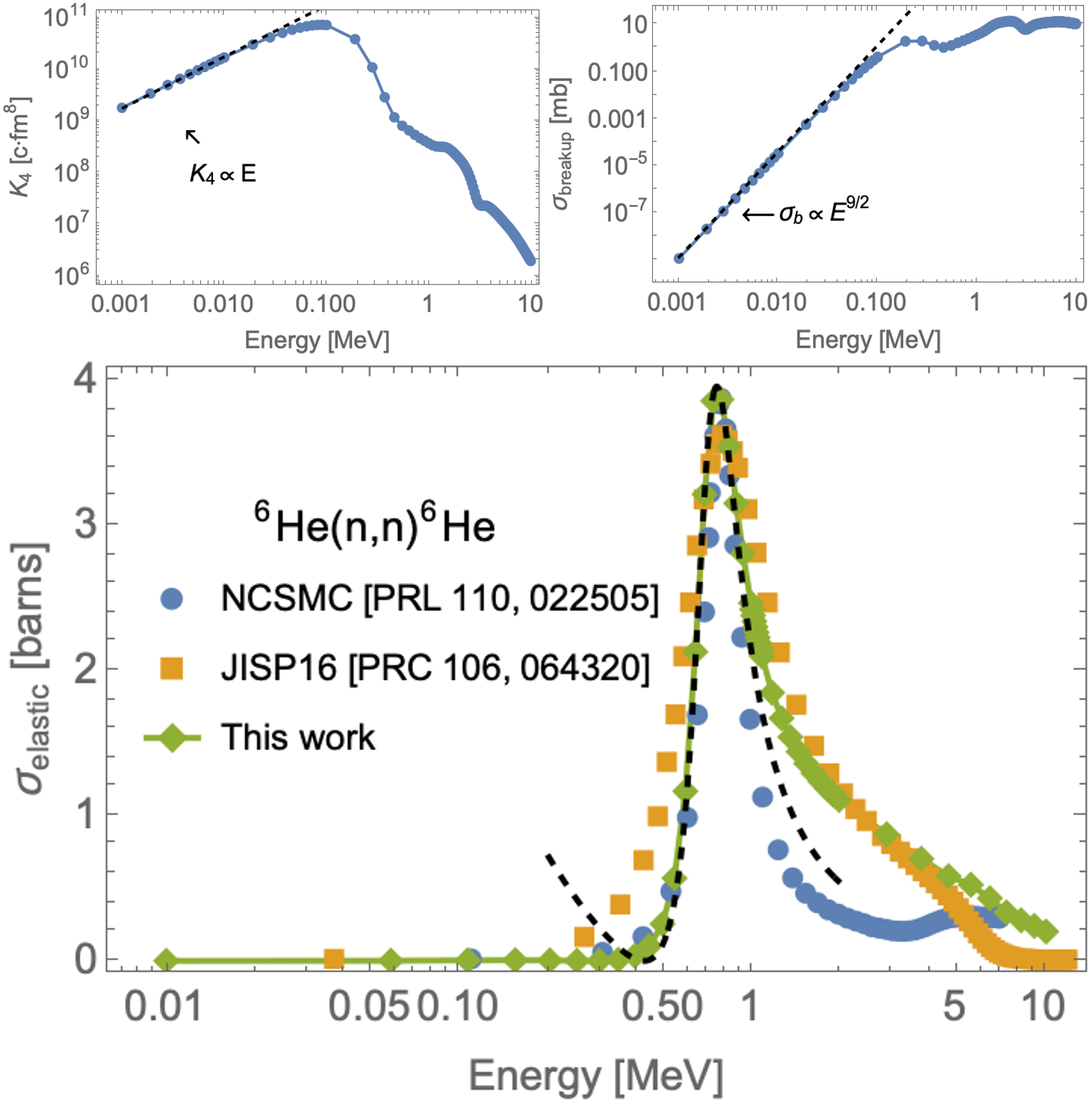}
\caption{\label{fig:he7_recombination} The four--body partial recombination rate (upper left panel) is shown for the process $n+n+n+\prescript{4}{}{\mathrm{He}}\rightarrow n+\prescript{6}{}{\mathrm{He}}(0^{+})$ in units of $c\cdot\mathrm{fm^{8}}$ for the symmetry $(L^{\pi},S)J^{\pi}=(1^{-},1/2)3/2^{-}$. The partial breakup cross section (upper right panel) is shown for the process $n+\prescript{6}{}{\mathrm{He}}(0^{+})\rightarrow n+n+n+\prescript{4}{}{\mathrm{He}}$ in units of $mb$. The energy axis on the plots in the upper panels are relative to the four--body breakup threshold. At low energies, the recombination rate follows the typical threshold law behavior, $K_{4}\propto k^{2l+1-(3N-5)}$, while the breakup cross section goes as $\sigma_{b}\propto k^{2l+1}$. For this system $l=4$ is the effective angular momentum quantum number in the lowest entrance channel, so the threshold law behavior of the recombination rate goes as $k^2$ or $E$ and the breakup cross section goes as $k^9$ or $E^{9/2}$. This behavior is highlighted by the dashed lines in the upper panels. The elastic $\prescript{6}{}{\mathrm{He}}+n$ cross section (lower panel) is shown as function of energy relative to the two--body breakup threshold. The peak in the cross section corresponds to the $\prescript{7}{}{\mathrm{He}}$ resonance. Near the resonance, a Breit--Wigner distribution is used to describe the peak, as shown by the dashed line. Previous theoretical calculations using the no--core shell model with continuum states \cite{PhysRevLett.110.022505,PhysRevC.87.034326} and \cite{PhysRevC.106.064320} are compared with this work, shown in the circles and squares, respectively.}
\end{figure}

Solving the coupled hyperradial Shr$\mathrm{\ddot{o}}$dinger equation for energies above the $n+n+n+\prescript{4}{}{\mathrm{He}}$ threshold ($E>0$) gives quantitative insights into four--body recombination and breakup processes in this system, namely the process $n+n+n+\prescript{4}{}{\mathrm{He}}\rightarrow n+\prescript{6}{}{\mathrm{He}}$ and the reverse process, respectively. The $N$--body recombination rate \cite{Rittenhouse_2011} to recombine in the $i^{\mathrm{th}}$ channel in terms of the $S$--matrix is given by Eq. \eqref{eq:nbody_recomb}
\begin{equation}
K_{N,i}^{J^{\pi}}=\frac{\hbar}{\mu}\frac{2\pi N_{p}}{\Omega(3N-3)}\left(\frac{2\pi}{k}\right)^{3N-5}\sum_{j\neq i}|S_{ij}|^2
\label{eq:nbody_recomb}
\end{equation}
where $k$ is the hyperradial wave number, $N_{p}$ is the number of identical particle permutations, and $|S_{ij}|^2$ represents the probability of scattering into channel $j$ from an initial channel $i$. The break--up cross section for the process $AB^2+B\rightarrow A+B+B+B$ is given by Eq. \eqref{eq:breakup_crosssection} as
\begin{equation}
    \sigma_{b}=\frac{\pi}{k_{i}^{2}}\sum_{f}|S_{if}|^2
    \label{eq:breakup_crosssection}
\end{equation}
where $k_{i}$ is the two--body wave number associated with the $AB^{2}+B$ system. Figure \ref{fig:he7_recombination} shows the partial recombination rate (see the upper left panel) as a function of energy relative to the four--body break--up threshold. In Fig. \ref{fig:he7_recombination}, the four--body recombination rate exhibits a threshold dependence that scales linearly with the four--body center of mass collision energy $E$, namely $K_{4}\propto E$. This linear behavior is shown by the black dashed line and agrees well with the calculated recombination rate for energies in the range $1~\mathrm{keV}<E<100~\mathrm{keV}$. The recombination rate has a maximum value on the order of $10^{11}~c\cdot\mathrm{fm}^{8}$ in the energy range of $100~\mathrm{keV}-200~\mathrm{keV}$. Figure \ref{fig:he7_recombination} also shows the partial break--up cross section (upper right panel) and the partial elastic cross section (lower panel) for this symmetry. The breakup cross section is suppressed at low collision energies and has a scaling law behavior of $\sigma_{\mathrm{break-up}}\propto E^{9/2}$. This suppression is due to the tunneling of the wave function under the angular momentum barrier in the lowest entrance channel. The angular momentum quantum number of the lowest entrance channel is $l=4$, giving rise to the suppression of the cross section as $E^{9/2}$ from the $k^{2l+1}$ threshold law behavior of $|S_{if}|^{2}$.
%\begin{figure}[ht!]
%\centering
%\includegraphics[width=1.0\columnwidth]{combined_graphic2.pdf}
%\includegraphics[width=8.6 cm]{he7_K4_Channels_3.eps}
%\caption{The four--body partial recombination rate is shown for the process $n+n+n+\prescript{4}{}{\mathrm{He}}\rightarrow n+\prescript{6}{}{\mathrm{He}}(0^{+})$ in units of $c\cdot\mathrm{fm^{8}}$ for the symmetry $(L^{\pi},S)J^{\pi}=(1^{-},1/2)3/2^{-}$. At low energies, the recombination rate follows the typical threshold law behavior which goes as $k^{2l+1-(3N-5)}$, while the cross section follows the typical threshold law behavior which goes as $k^{2l+1}$, where $l$ is the effective angular momentum quantum number in the lowest entrance channel. For this system, $N=4$ and $l=4$, so the threshold law behavior of the recombination rate goes as $k^2$ or $E$. This behavior is highlighted by the dashed line on the figure for low energies relative to the four--body continuum threshold.}
%\label{fig:he7_recombination}
%\end{figure}

In an experiment, the elastic scattering cross section \cite{Rittenhouse_2011} for the $\prescript{6}{}{\mathrm{He}}(n,n)\prescript{6}{}{\mathrm{He}}$ process could be used to measure the $\prescript{7}{}{\mathrm{He}}$ resonance. The $N$--body elastic cross section is given by Eq. \eqref{eq:elastic_cross}. For the $\prescript{6}{}{\mathrm{He}}(n,n)\prescript{6}{}{\mathrm{He}}$ process, $N=2$ and the cross section has units of $\mathrm{fm}^{2}$, or barns.
\begin{equation}
\label{eq:elastic_cross}
\sigma_{\mathrm{elastic},i}=\frac{1}{\Omega(3N-3)}\left(\frac{2\pi}{k}\right)^{3N-4}|S_{ii}-1|^{2}
\end{equation}
The elastic cross section for this process in the $(1^{-},1/2)3/2^{-}$ symmetry is shown in the lower panel of Fig. \ref{fig:he7_recombination}. There is a peak in the cross section that corresponds to the $\prescript{7}{}{\mathrm{He}}$ resonance. The elastic cross section near the resonance peak follows a Breit--Wigner distribution. The form of the Breit--Wigner distribution is given in Eq. \eqref{eq:breit_wigner}
\begin{multline}
\label{eq:breit_wigner}
\sigma_{bw}=\frac{4\pi}{k^2}\mathrm{sin}^{2}(\delta_0)\left[\frac{(\Gamma/2)^{2}}{(\Gamma/2)^{2}+(E-E_{r})^{2}}\right]\\\times\left(1+\frac{2(E-E_{r})\mathrm{cot}(\delta_{0})}{\Gamma}\right)^{2}
\end{multline}
where $\delta_{0}$ is the phase shift at the resonance energy $E_{r}$, and $\Gamma$ is the resonance width \cite{PhysRev.49.519}. This formula can be derived by performing a Taylor series expansion of Eq. \eqref{eq:elastic_cross} about $E=E_{r}$ for $N=2$. The Breit--Wigner distribution is shown as the dashed line in the lower panel of Fig. \ref{fig:he7_recombination} using the resonance parameters given in Table \ref{table:three_body_phase_table_he7} for the 8--channel calculation. Also shown in the lower panel of Fig. \ref{fig:he7_recombination} are two different NCSM calculations \cite{PhysRevLett.110.022505,PhysRevC.87.034326,PhysRevC.106.064320}. The cross sections were extracted from the energy--dependent phase shifts plotted in these references. The calculations in \cite{PhysRevLett.110.022505,PhysRevC.87.034326} were performed using the chiral $\mathrm{N}^{3}\mathrm{LO}$ $NN$ interaction, giving a resonance and width of $0.71~\mathrm{MeV}$ and $0.30~\mathrm{MeV}$, respectively. The calculations in \cite{PhysRevC.106.064320} were performed using the JISP16 $NN$ interaction \cite{PhysRevC.70.044005}, giving a resonance and width of $0.665~\mathrm{MeV}$ and $0.57~\mathrm{MeV}$, respectively. These two calculations show both qualitative and quantitative agreement with the results obtained in this work, demonstrating the accuracy of the model and method used and how it compares to commonly used methods in nuclear physics. The NCSM calculations with realistic $NN$ interactions yield similar resonance positions and widths for the $3/2^{-}$ symmetry to the ones computed in this work. The differences among the theory calculations can be attributed to the different nuclear Hamiltonians used and the methods implemented.

The model Hamiltonian constructed reproduces the ground state and resonance state of the $\prescript{6}{}{\mathrm{He}}$ nucleus, as well as describes a resonance in the $(L^{\pi},S)J^{\pi}=(1^{-},1/2)3/2^{-}$ symmetry for the $\prescript{7}{}{\mathrm{He}}$ nucleus. Thus, the next step is to study the $\prescript{4}{}{\mathrm{He}}+4n$ system to see how well the Hamiltonian describes the states of the $\prescript{8}{}{\mathrm{He}}$ nucleus.

\subsection{Some Properties of the \texorpdfstring{$\prescript{8}{}{\mathrm{He}}$}{text} Nucleus}
\label{sec:five_body_sys}
From a full diagonalization of the Hamiltonian, the $\prescript{8}{}{\mathrm{He}}(0^{+})$ ground state energy was computed and compared with experiment. These results are shown in Table \ref{table:five_body_J0_table}. The parameters of the spin--triplet three--body interaction was fit to reproduce the ground state energy, rms matter radius, and the rms distance of the $\prescript{4}{}{\mathrm{He}}$ nucleus from the four--neutron center of mass. These results are shown in the first row of Table \ref{table:five_body_J0_table}. The ground state energy is -3.114 MeV, which agrees well with the experimental value of -3.11 MeV \cite{PhysRevLett.101.012501} and only differs by 0.13$\%$. The rms matter radius was computed to be 2.54 fm, which compared to the experimental value of 2.52(3) fm \cite{WAKASA2022105329}, only differs by 0.8$\%$. The rms value of the distance of the $\prescript{4}{}{\mathrm{He}}$ nucleus from the four neutron center of mass is computed to be 2.381 fm. The rms matter radius and the rms distance between the $\prescript{4}{}{\mathrm{He}}$ nucleus and the four--neutron center of mass was computed assuming an rms matter radius of 1.57 fm for the $\prescript{4}{}{\mathrm{He}}$ nucleus \cite{TANIHATA1992261,WAKASA2022105329}. 

The model Hamiltonian used reproduces the ground--state properties of the $\prescript{8}{}{\mathrm{He}}$ nucleus. As previously stated, the model Hamiltonian produces a $\prescript{7}{}{\mathrm{He}}(3/2^{-})$ resonance that is higher in energy than the experimental value (see Table \ref{table:three_body_phase_table_he7}). If the strength of the spin--triplet three--body interaction is increase by a factor of approximately 1.7, the $\prescript{7}{}{\mathrm{He}}(3/2^{-})$ resonance and width are shifted to the experimental values to within a few percent. However, this leads to the $\prescript{8}{}{\mathrm{He}}(0^{+})$ ground--state becoming overbound by approximately $22\%$ from the experimental value, whereas the rms matter radius and $\prescript{4}{}{\mathrm{He}}-4n$ separation differ by approximately $2\%$. These shifted values are shown in the second row of Table \ref{table:five_body_J0_table}.

\begin{table}[H]
\caption{The $\prescript{8}{}{\mathrm{He}}$ $0^{+}$ ground--state energy $E_{\mathrm{g.s.}}$, the rms matter radius $\langle r^{2}\rangle^{\frac{1}{2}}_{\mathrm{rms}}$, and the $\prescript{4}{}{\mathrm{He}}-4n$ rms separation $\langle r^{2}\rangle^{\frac{1}{2}}_{\mathrm{\alpha-4n}}$ are tabulated for different values of the spin--triplet three--body interaction strength, denoted $V_{1}^{(3b)}$. The symbol $\alpha$ is used to label the $\prescript{4}{}{\mathrm{He}}$ nucleus. These calculations were performed via a full diagonalization of the five--body Hamiltonian using an ECG basis. The rms matter radius for the $\prescript{4}{}{\mathrm{He}}$ nucleus is taken to be 1.57~fm \cite{TANIHATA1992261,WAKASA2022105329}. The values in square brackets are taken from \cite{PhysRevLett.101.012501,TANIHATA1992261,WAKASA2022105329}.}
\label{table:five_body_J0_table}
\begin{ruledtabular}
\begin{tabular}{cccc}
    $V_{1}^{(3b)}~[\mathrm{MeV}]$ & $E_{\mathrm{g.s.}}~[\mathrm{MeV}]$ & $\langle r^{2}\rangle^{\frac{1}{2}}_\mathrm{rms}~[\mathrm{fm}]$ & $\langle r^{2}\rangle^{\frac{1}{2}}_\mathrm{\alpha-4n}~[\mathrm{fm}]$\\
    \hline
    $-5.08$ & $-3.114~[-3.11]$ & $2.540~[2.52(3)]$ & $2.381~[2.367]$\\
    %$-3.371^{*}~[-3.11]$ & $2.512^{*}~[2.52(3)]$ & $2.351^{*}~[2.367]$\\
    $-7.00$ & $-3.80~[-3.11]$ & $2.47~[2.52(3)]$ & $2.31~[2.367]$\\
\end{tabular}
\end{ruledtabular}
\end{table}

\begin{figure}[ht!]
% Center the entire figure (containing the two subfigures).
\centering
\includegraphics[width=1.0\columnwidth]{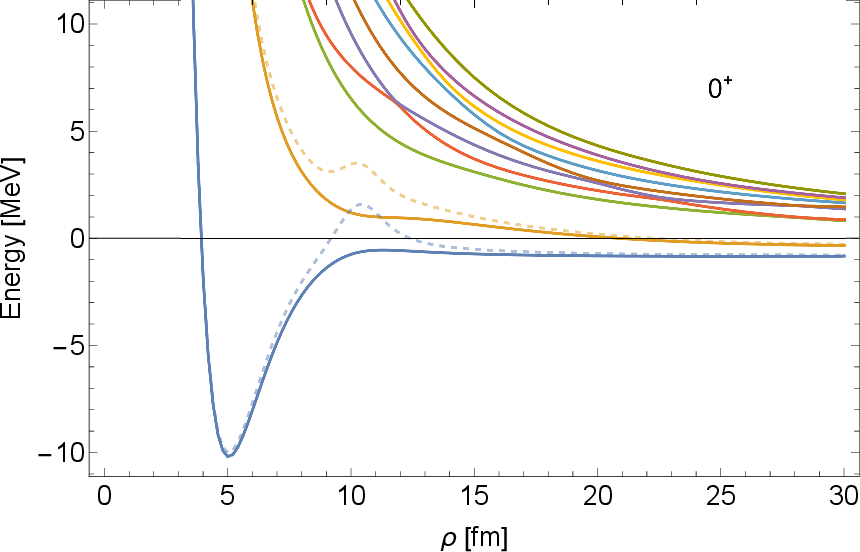}
%\subfigure[]{\includegraphics[width=8.6 cm]{He8_J_0p.eps}}\label{fig:he8_pot_a}
%\subfigure[]{\includegraphics[width=8.6 cm]{He8_2p.eps}}\label{fig:he8_pot_b}
\caption{\label{fig:helium8_potentials}The lowest few adiabatic potential curves for the $\prescript{4}{}{\mathrm{He}}+4n$ system in the $(L^{\pi},S)J^{\pi}=(0,0)0^{+}$ symmetry. The solid curves are the adiabatic potentials and the dashed curves are the effective potentials, which includes the diagonal second--derivative coupling term. Only the lowest two effective potentials are shown here.}
\end{figure}

With two-- and three--body interactions between an $\prescript{4}{}{\mathrm{He}}$ nucleus and neutrons describing the lowest $\prescript{6}{}{\mathrm{He}}$ and $\prescript{7}{}{\mathrm{He}}$ bound, resonant, as well as continuum states, efforts have been made to utilise the adiabatic hyperspherical approach to solve the five--body problem to describe the $\prescript{8}{}{\mathrm{He}}$ nucleus. An unconverged calculation of the lowest few hyperradial potential curves for the $\prescript{8}{}{\mathrm{He}}$ system in the $0^{+}$ symmetry are shown in Figure \ref{fig:helium8_potentials}. In principle, a fully converged calculation would show an infinite number of hyperradial potential curves that crash down and approach the $\prescript{6}{}{\mathrm{He}}(0^{+})+n+n$ threshold, which lies $0.98~\mathrm{MeV}$ below the five--body continuum. From Fig. \ref{fig:helium8_potentials}, only the lowest two potential curves go asymptotically to this threshold. More computational effort is required to get higher channels to converge enough to accurately perform multichannel scattering calculations to study processes like five--body recombination ($\prescript{4}{}{\mathrm{He}}+n+n+n+n\rightarrow\prescript{6}{}{\mathrm{He}}+n+n$) and the breakup process ($\prescript{6}{}{\mathrm{He}}+n+n\rightarrow\prescript{4}{}{\mathrm{He}}+n+n+n+n$).

\section{Summary}
\label{sec:conclusion}
A simplified model of the $\prescript{4}{}{\mathrm{He}}-n$ and $\prescript{4}{}{\mathrm{He}}-nn$ systems has been implemented in the adiabatic hyperspherical approach to study how few--neutron clusters interact in the presence of an $\prescript{4}{}{\mathrm{He}}$ nucleus.

First, with the simplified angular momentum-dependent interaction for the $\prescript{4}{}{\mathrm{He}}-n$ two--body system, the three--body $\prescript{6}{}{\mathrm{He}}$ system was studied. A spin--dependent three--body $\prescript{4}{}{\mathrm{He}}-nn$ interaction was introduced into the Hamiltonian. The parameters for the spin--singlet interaction were tuned to give the correct $\prescript{6}{}{\mathrm{He}}(0^{+})$ ground--state energy and rms matter radius, as well as the $\prescript{6}{}{\mathrm{He}}(2^{+})$ resonance and width. The parameters for the spin--triplet interaction were tuned to reproduce the $\prescript{8}{}{\mathrm{He}}(0^{+})$ ground--state energy, rms matter radius, and the rms distance from the $\prescript{4}{}{\mathrm{He}}$ nucleus to the $4n$ center of mass. 

With these two-- and three--body interactions, $\prescript{6}{}{\mathrm{He}}$ and $\prescript{7}{}{\mathrm{He}}$ systems were investigated to characterize bound-- and resonance states, along with computing elastic and inelastic collision processes. For the $\prescript{6}{}{\mathrm{He}}$ system, the electric quadrupole transition matrix element was computed between the $0^{+}$ ground--state and $2^{+}$ three--body continuum states as a function of the three--body energy relative to the center of mass. The functional behavior of the quadrupole matrix element is in good qualitative agreement with other theoretical calculations in the literature. There is a clear peak in the quadrupole transition at an energy around the $\prescript{6}{}{\mathrm{He}}(2^{+})$ resonance energy, to within the width.

The $\prescript{7}{}{\mathrm{He}}$ system was also studied, treated as a four--body system of an $\prescript{4}{}{\mathrm{He}}$ nucleus interacting with three neutrons. The symmetry studied in this work was the $3/2^{-}$ symmetry, which has a known resonance. The simplified model explored in this work supports a $3/2^{-}$ resonance, with its energy and width given in Table \ref{table:three_body_phase_table_he7}. The calculated energy and width are larger than the experimental values. Enhancing the strength of the spin--triplet three--body interaction by a factor of approximately 1.7 shifts the resonance and width to within a few percent of the experimental values, but shifts the $\prescript{8}{}{\mathrm{He}}(0^{+})$ ground state energy away from its experimental value by more than $20\%$.

Scattering processes for the $\prescript{7}{}{\mathrm{He}}$ four--body system were investigated. These processes include four--body recombination ($\prescript{4}{}{\mathrm{He}}+n+n+n\rightarrow\prescript{6}{}{\mathrm{He}}+n$), two--body breakup ($\prescript{6}{}{\mathrm{He}}+n\rightarrow\prescript{4}{}{\mathrm{He}}+n+n+n$), and two--body elastic scattering ($\prescript{6}{}{\mathrm{He}}+n\rightarrow\prescript{6}{}{\mathrm{He}}+n$). The four--body recombination process, which describes the formation of $\prescript{6}{}{\mathrm{He}}$ in its ground state through the collision of $\prescript{4}{}{\mathrm{He}}$ with two free neutrons is computed for energies ranging from $1~\mathrm{keV}$ to $10~\mathrm{MeV}$. At low energies the recombination rate scales as $K_{4}\propto E$ and has a maximum value of approximately $10^{11}~c\cdot \mathrm{fm}^{8}$ in the energy range $100~\mathrm{keV}-200~\mathrm{keV}$. The computed elastic $\prescript{6}{}{\mathrm{He}}+n$ cross section is on the order of a few barns and exhibits a peak associated with the $\prescript{7}{}{\mathrm{He}}(3/2^{-})$ resonance that follows the Breit--Wigner distribution. The computed $\prescript{6}{}{\mathrm{He}}+n$ breakup cross section is on the order of a few millibarns. The low--energy threshold behavior for the breakup cross section follows the Wigner threshold law and leads to the scaling $\sigma_{\mathrm{el}}\propto E^{9/2}$.

\section{Acknowledgements}
\noindent This work is supported
in part by the NSF, Grant
No. PHY-2207977, and in part by the Purdue Quantum
Science and Engineering Institute.

%\bibliography{all-biblatex}
\bibliography{bibliography}
\end{document}